\documentclass[11pt]{article}
\usepackage{moriond,epsfig}

\def\Journal#1#2#3#4{{#1} {\bf #2}, #3 (#4)}

\def\NPB{{\em Nucl. Phys.} B}
\def\PLB{{\em Phys. Lett.}  B}

\def\PRD{{\em Phys. Rev.} D}

\def\be{\begin{equation}}
\def\ee{\end{equation}}
\def\bea{\begin{eqnarray}}
\def\eea{\end{eqnarray}}

\begin{document}
\vspace*{4cm}
\title{H1 AND ZEUS RESULTS ON BEAUTY PRODUCTION}

\author{ I.-A. MELZER-PELLMANN }

\address{DESY, ZEUS Group, Notkestr. 85,\\
D-22607 Hamburg, Germany}

\maketitle\abstracts{
The H1 and ZEUS experiments are measuring beauty
production in ep collisions at HERA.
The $b$ quark mass provides a hard scale which allows
a dedicated test of perturbative QCD.
The latest results in photoproduction and deep inelastic
scattering are presented and compared to the NLO QCD 
predictions as well as to LO + parton shower Monte Carlo simulations.}

\section{Introduction} \label{Melzer:Introduction}
At HERA a center of mass energy of $\sqrt{s}=318$ GeV is achieved by colliding
positrons (or electrons) of 27.5 GeV and protons of 920 GeV. In the years 
1996-2000 the experiments H1 and ZEUS collected an integrated luminosity of 
more than 100~pb$^{-1}$ each, yielding a good opportunity to study perturbative
QCD calculations for beauty production.

Photon gluon fusion is the main mechanism for beauty production at HERA, 
where a photon emitted by the initial positron~\footnote{Electrons and positrons will 
not be distinguished in the following sections and are both referred to as positrons.}
interacts with a gluon in 
the proton, resulting in a quark antiquark pair. With at least one hard scale, due to 
the large $b$ quark mass, perturbative QCD calculations are applicable. Nevertheless, 
a hard scale is also given by the transverse jet energy and, in DIS events, by the photon 
virtuality $Q^2$. The presence of more than one hard scale can lead to large
logarithms in the calculation, which can spoil the reliability of the perturbative
expansion. Hence, precise differential cross section measurements are needed
to test the theoretical understanding of beauty production.

The measurement techniques rely on signatures that are characteristic for $b$ production
and decay. Due to the large $b$ mass, the transverse momentum of the muon with respect to
the axis of the closest jet, $p_T^{rel,\mu}$, is larger in a $b$ decay to a jet and a muon 
than in decays of lighter quarks. A different method to extract $b$ decays is the measurement
of the lifetime provided by silicon vertex detectors. Because of the large $b$ lifetime, the 
impact parameter of the muon is larger for $b$ than for light flavour decays.

\section{Beauty in Photoproduction} \label{Melzer:PHP}

\begin{figure} [t]
  \begin{center}
    \begin{tabular}{cc}
      \mbox{\epsfig{figure=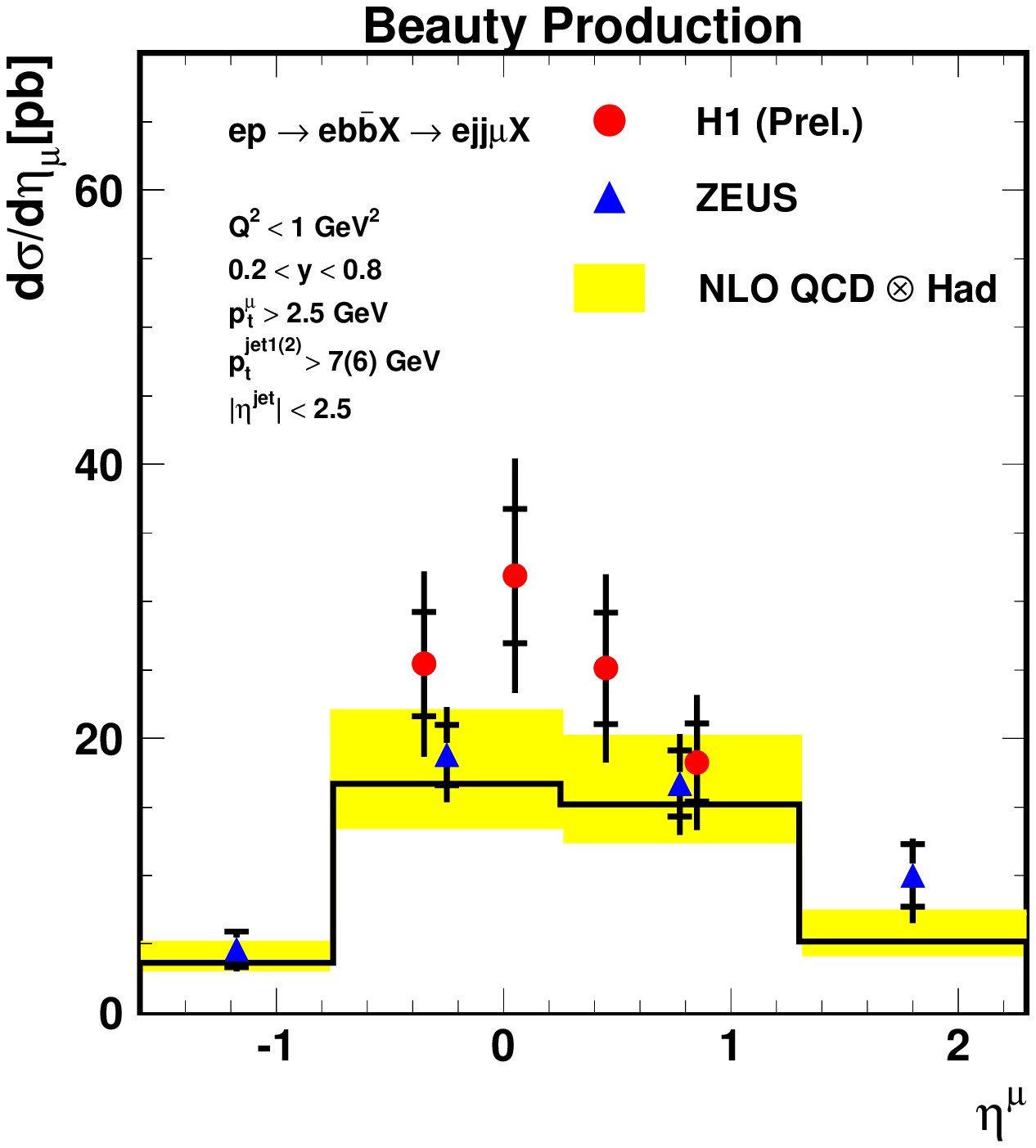,width=0.4\textwidth}} &
      \mbox{\epsfig{figure=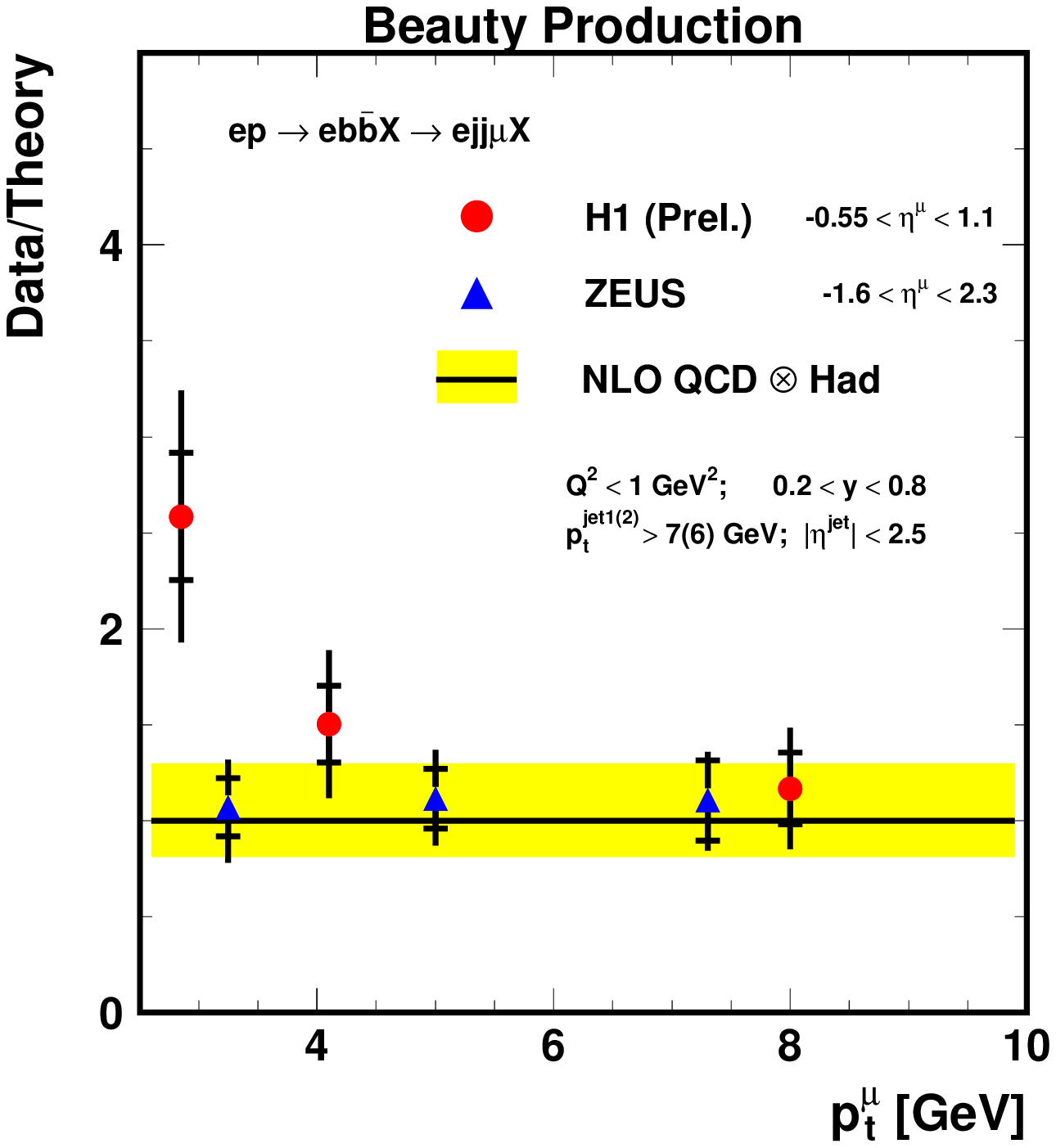,width=0.4\textwidth}}
    \end{tabular}
  \end{center}
  \caption{Differential cross sections of beauty photoproduction as a function
    of the muon pseudorapidity in comparison to NLO QCD (left) and the data/theory 
    ratio of the muon transverse momentum (right).
    \label{Melzer:fig:php}}
\end{figure}

If the negative squared four-momentum exchanged at the positron vertex
is small ($Q^2 < 1$ GeV$^2$), the exchanged photon emitted by the positron is quasi-real,
and the reaction $ep \rightarrow e' b \bar{b} X$ can be considered as a photoproduction 
process.

ZEUS has measured beauty photoproduction~\cite{Z-PHP} in events with two jets and a muon
in the final state, using the data collected in the years 1996-1997 and 1999-2000, which
correspond to an integrated luminosity of $\mathcal{L} = 110.4 \pm 2.2$~pb$^{-1}$. The 
visible cross sections are obtained for events with $p_T^{jet_{1(2)}} > 7(6)$ GeV,
$\eta^{jet_{1,2}} < 2.5$, $0.2 < y < 0.8$ and $Q^2 < 1$ GeV$^2$. 
The angular acceptance of the three muon chambers and of the central tracking detector 
define three regions of good acceptance, in which the cross section is measured and compared to 
the NLO QCD prediction based on the FMNR program~\cite{FMNR} with Peterson fragmentation~{\cite{Pet83} 
(see Tab.~\ref{tab:z-php}):
\begin{table}[!h]
\caption{ZEUS kinematic ranges and cross sections for each muon chamber in comparison 
to the NLO QCD prediction corrected to the hadron level with 
the theoretical uncertainty and the hadronisation corrections $C_{had}$.\label{tab:z-php}}
\vspace{0.4cm}
\begin{center}
\begin{tabular}{|r|c|c|c|l|}
\hline
$\mu$-chambers & kinematic range & $\sigma \pm$ stat. $\pm$ sys. & $\sigma^{NLO} \times C_{had}$ & $C_{had}$\\
\hline
rear & $ -1.6 < \eta^\mu < -0.9$, $p^\mu > 2.5$ GeV  & $6.5 \pm 1.5 ^{+1.0}_{-1.1} $ & $4.3^{+1.6}_{-1.0} $  & 0.80 \\
barrel & $ -0.9 < \eta^\mu < 1.3$, $p^\mu_T > 2.5$ GeV & $38.2 \pm 3.4 ^{+5.7}_{-5.8}$ & $33.9 ^{+11.0}_{-7.0}$ & 0.89 \\
forward & $ 1.48 < \eta^\mu < 2.3$, $p^\mu > 4$ GeV,  & $16.6 \pm 3.3 ^{+2.9}_{-4.6}$ & $6.5^{+2.8}_{-1.6} $ & 0.86 \\
 & $p^\mu_T > 1$ GeV &  &  & \\
\hline
\end{tabular}
\end{center}
\end{table}

H1 has used data collected in the years 1999-2000 corresponding to an integrated luminosity 
of $\mathcal{L} = 48$~pb$^{-1}$ to measure beauty photoproduction~\cite{H-PHP}. The visible cross
section for events with at least two high transverse momentum jets with $p_T^{jet_{1(2)}} > 7(6)$ GeV,
$\eta^{jet_{1,2}} < 2.5$, $0.2 < y < 0.8$, $Q^2 < 1$ GeV$^2$, and a muon in the final state with 
$ -0.55 < \eta^\mu < 1.1$ and $p^\mu_T > 2.5$ GeV is measured to be
$\sigma_{vis}(ep \rightarrow e b \bar{b} X \rightarrow e j j \mu X) = 42.5 \pm 3.4_{stat} \pm 8.9_{sys}$.\\
The prediction from the NLO QCD calculation~\cite{FMNR} including fragmentation~\cite{Pet83} and hadronisation 
corrections is 
$\sigma^{NLO} \times C_{had} = 24.1^{+7.2}_{-5.1}$. 

The differential cross section in $\eta^\mu$ and the data/theory ratio as a function of $p^\mu_T$
measured by ZEUS and H1 are shown in Fig.~\ref{Melzer:fig:php}.

\section{Beauty in Deep Inelatic Scattering} \label{Melzer:DIS}

\begin{figure} [t]
  \begin{center}
    \psfig{figure=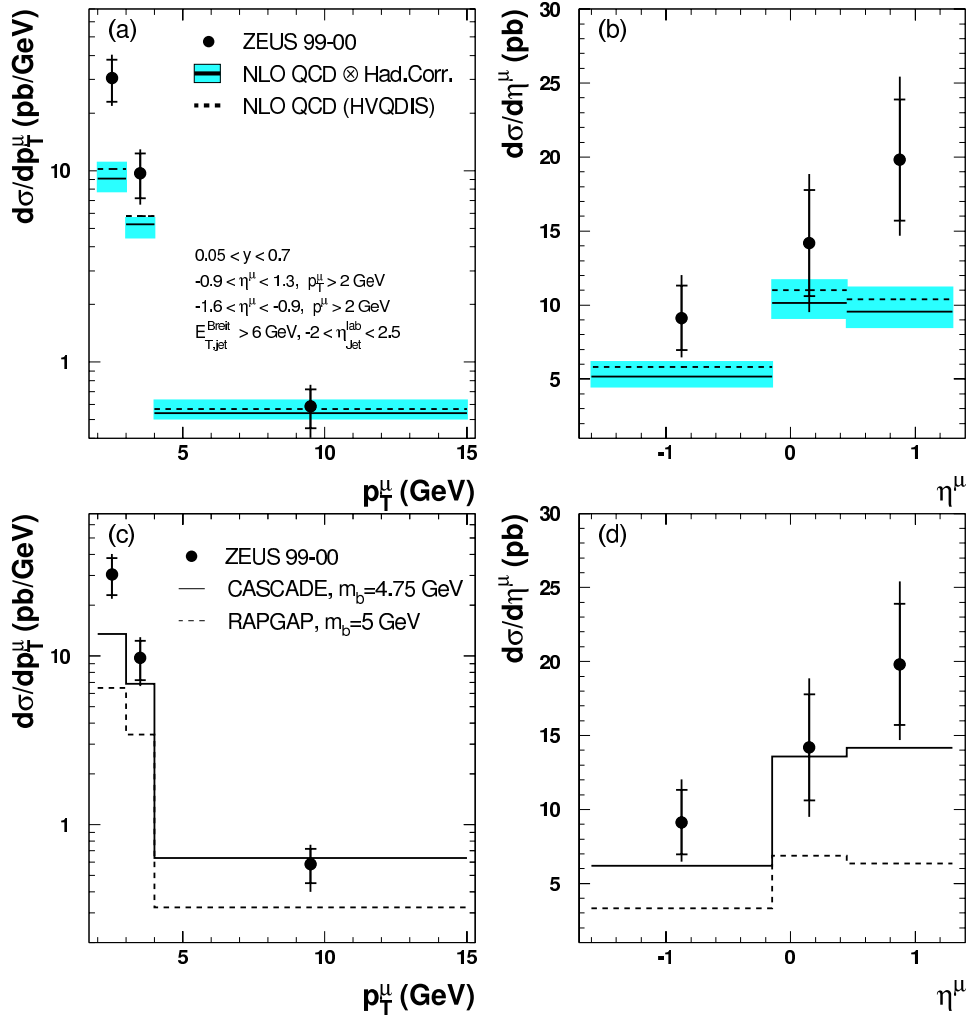,width=0.8\textwidth}
  \end{center}
  \caption{Differential cross sections of beauty in DIS measured by ZEUS as a function
    of the muon transverse momentum (left) and of its pseudorapidity (right)
    in comparison to NLO QCD calculations (a,b) and to the LO QCD MC programs (c,d)
    {\sc Cascade} (solid line) and {\sc Rapgap} (dashed line).
    \label{Melzer:fig:diszeus}}
\end{figure}

ZEUS has measured beauty production in DIS~\cite{Z-DIS} using the data taken in 1999-2000, which
corresponds to an integrated luminosity of $\mathcal{L} = 72.4 \pm 1.6$~pb$^{-1}$. The
total visible cross section was determined in the kinematic range $Q^2 >2$ GeV$^2$ and $0.05 < y < 0.7$
with at least one hadron-level jet in the Breit frame~\footnote{The Breit frame is defined by
$\vec{\gamma}+2x\vec{P}=\vec{0}$, where $\vec{\gamma}$ is the momentum of the exchanged photon,
$x$ the Bjorken scaling variable and $\vec{P}$ is the proton momentum. A space-like photon and
a proton collide head-on.} with $E_{T,jet}^{Breit} > 6$ GeV and $-2 < \eta^{lab}_{jet} < 2.5$, and 
with a muon fulfilling: $ -1.6  < \eta^\mu < -0.9$ and $p^\mu > 2$ GeV or 
$ -0.9 < \eta^\mu < 1.3$ and $p^\mu_T > 2$ GeV. The result is 
$\sigma_{vis}(ep \rightarrow e b \bar{b} X \rightarrow e j \mu X) = 40.9 \pm 5.7_{stat}\,^{+6.0}_{-4.4\, sys}$.\\
The NLO QCD calculation using the {\sc Hvqdis} program~\cite{Har} with hadronisation corrections modeled
by the Kartvelishvili~\cite{Kar} parametrisation predicts $20.6 ^{+3.1}_{-2.2}$~pb, while the 
{\sc Cascade} Monte Carlo program gives 28~pb and the {\sc Rapgap} Monte Carlo gives 14~pb. 
Fig.~\ref{Melzer:fig:diszeus} shows that the data are well described by NLO QCD and by the MC except
in the region of low $p^\mu_T$.

\begin{figure} [t]
  \begin{center}
    \begin{tabular}{cc}
      \mbox{\epsfig{figure=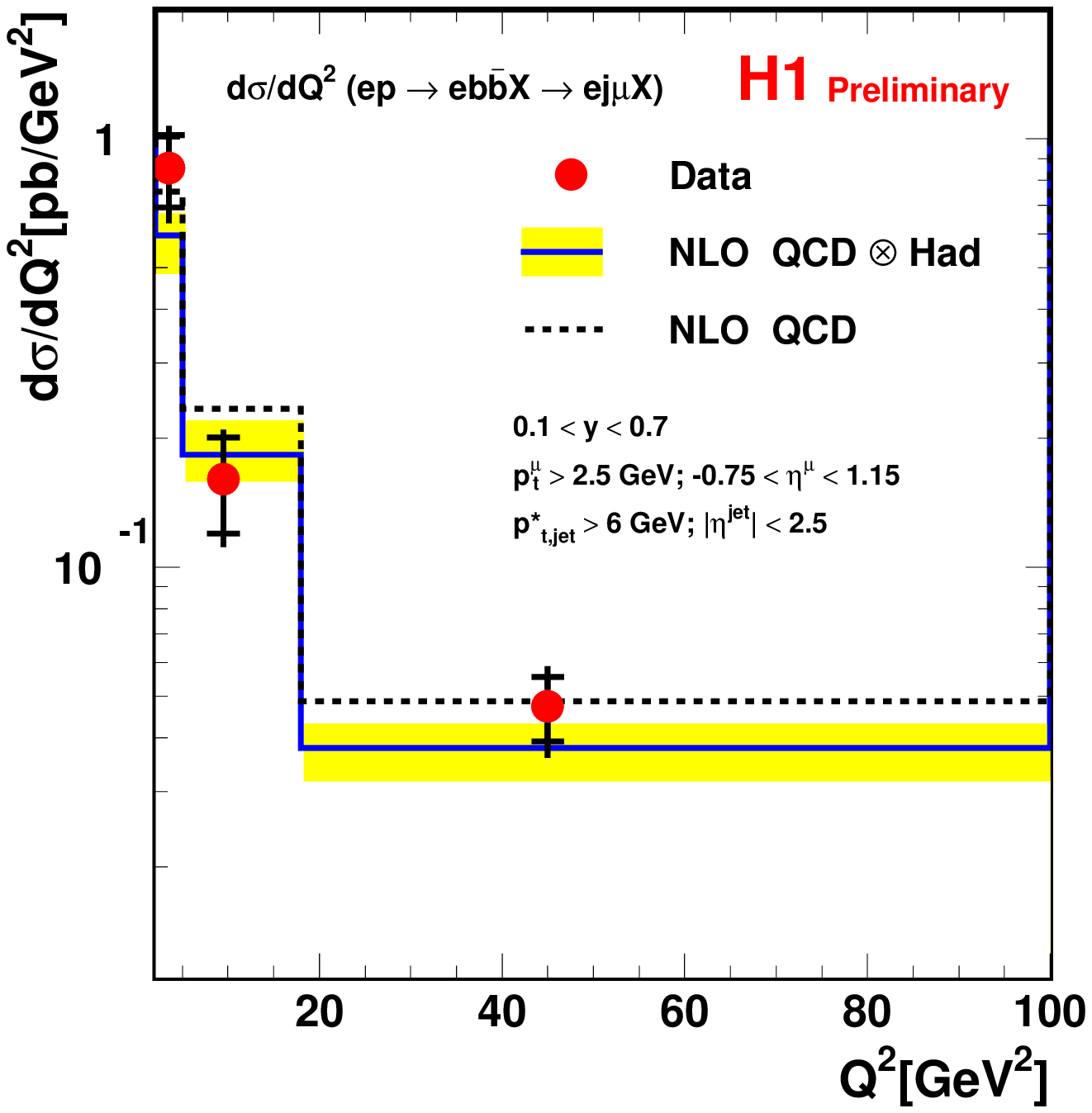,width=0.4\textwidth}} &
      \mbox{\epsfig{figure=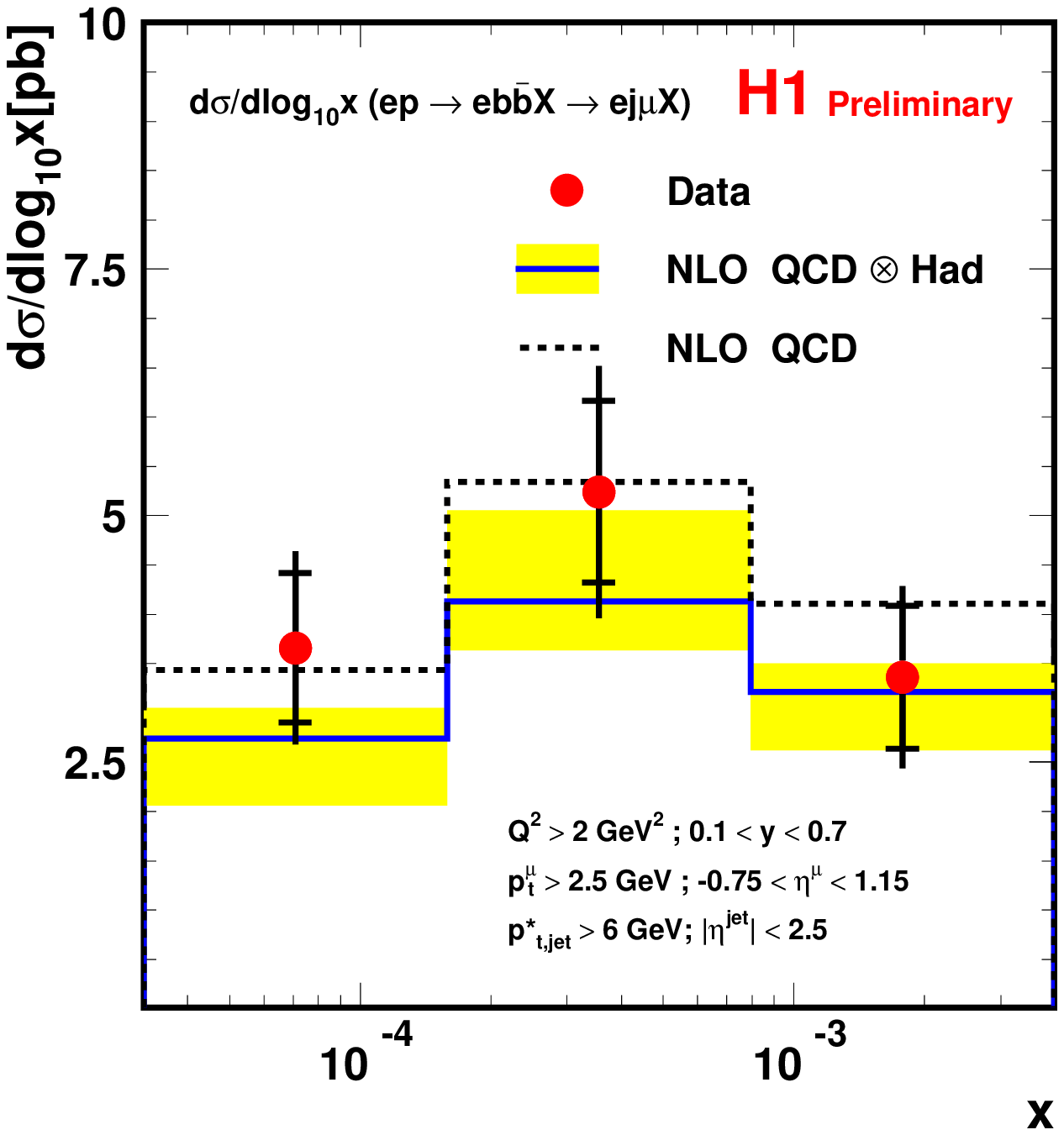,width=0.4\textwidth}}
    \end{tabular}
  \end{center}
  \caption{Differential cross sections of beauty production in DIS measured by H1 as a function
    of $Q^2$ (left) and of $x$ (right) in comparison to NLO QCD.
    \label{Melzer:fig:dish1}}
\end{figure}

H1 has also measured beauty production in DIS~\cite{H-DIS}, using the data collected in the years 1999-2000,
which correspond to an integrated luminosity of $\mathcal{L} = 50$~pb$^{-1}$. The events are selected by
requiring at least one high transverse momentum jet in the Breit frame with $p_{T,jet}^{Breit} > 6$ GeV,
$|\eta_{jet}| < 2.5$ and a muon associated to the jet with $-0.75 < \eta^\mu < 1.15$ and  $p^\mu_T > 2.5$ GeV.
The visible cross section in the range $2 < Q^2 < 100$ GeV$^2$ and $0.1 < y < 0.7$ is measured as
$\sigma_{vis}(ep \rightarrow e b \bar{b} X \rightarrow e j \mu X) = 8.8 \pm 1.0_{stat} \pm 1.5_{sys}$.\\
In comparison, the NLO QCD calculation, including corrections from fragmentation and hadronisation, is
$ 7.3 ^{+1.0}_{-1.5}$~pb. The Monte Carlo programs {\sc Rapgap} and {\sc Cascade} give a good description 
of the shapes of the distributions observed in the data, though the {\sc Rapgap} prediction is too low in 
normalization.

\section{Summary and Outlook}\label{Melzer:Summary}

Recent HERA beauty measurements show improved agreement between data and NLO QCD calculations. The Monte 
Carlo simulations describe the shape very well, though sometimes the predictions are too low in 
normalisation. 
The higher statistics which will be collected in the HERA II running period in the next years will allow 
more precise differential measurements.

\end{document}